\documentclass[twocolumn,aps,pre,showpacs]{revtex4}
\usepackage[latin1]{inputenc}
\usepackage{graphicx}
\usepackage{latexsym}
\usepackage{amssymb}
\usepackage{amsmath}

\def\d{{\rm d}}
\def\ex{{\rm e}}
\def\u{{\bf u}}
\def\v{{\bf v}}
\def\x{{\bf x}}

\def\r{{\bf r}}
\def\z{{\bf z}}

\def\rv{r_{\rm v}}

\def\bnu{{\boldsymbol\nu}}

\def\bomega{{\boldsymbol\omega}}
\def\smalze{{\scriptscriptstyle{(0)}}}
\def\smalun{{\scriptscriptstyle{(1)}}}
\def\smaldu{{\scriptscriptstyle{(2)}}}
\def\smalqu{{\scriptscriptstyle{(4)}}}
\def\smaln{{\scriptscriptstyle{(n)}}}
\def\smalnp1{{\scriptscriptstyle{(n+1)}}}

\def\beq{\begin{equation}}
\def\eeq{\end{equation}}
\begin{document}

\title{Preferential concentration vs. clustering in inertial particle transport
by random velocity fields}
\author{Piero Olla}
\affiliation{ISAC-CNR and INFN, Sez. Cagliari, I--09042 Monserrato, Italy.}
\date{\today}

\begin{abstract}
The concept of preferential concentration in the transport of inertial particles
by random velocity fields is extended to account for the possibility of zero
correlation time and compressibility of the velocity field. It is shown
that, in the case of an uncorrelated in time random velocity field,
preferential concentration takes the form of a condition on the field
history leading to the current particle positions. This generalized
form of preferential concentration appears to be a necessary condition
for clustering in the uncorrelated in time case. The standard 
interpretation of preferential concentration is recovered 
considering local time averages of the velocity field. 
In the compressible case, preferential concentration 
occurs
in regions of negative divergence of the field. In the
incompressible case, it  occurs
in regions of simultaneously high strain and negative
field skewness.


\end{abstract}

\pacs{05.10.Gg,05.40.-a,46.65.+g}
\maketitle

\section{INTRODUCTION}
\label{sec1}
One of the most striking characteristics of inertial particle transport by
random velocity fields is clustering. This phenomenon 
occurs, but is not confined to, in turbulent flows \cite{fessler94};
clustering phenomena, in fact, were initially predicted
to occur with particles pushed by a 1D (one-dimensional) random force
field \cite{deutsch85}. 
The interesting point is that
an initial spatially homogeneous
distribution of inertial particles will develop clumps and voids, 
even if the flow is incompressible. 
Both experimental evidence \cite{fessler94} and numerical simulations
\cite{hogan01,boffetta04} confirm this effect. 
Spatial inhomogeneity 
of the random field statistics may contribute, but is not crucial to the process.

It is to be mentioned that clustering phenomena are thought to be important 
both in industrial flows \cite{crowe98}, in the atmosphere
(the problem of rain formation) \cite{vaillancourt00,shaw03}
and in the oceans (the problem of plankton dynamics, especially as
regards blooming) \cite{ruiz04,reigada02}. 

Over the years, a substantial theoretical effort has been directed 
to the analysis of clustering by random flows
\cite{elperin96,balkovsky01,nishikawa01,sigurgeirsson02,fung03,zaichik03,duncan05,bec08}.
As first noticed in \cite{deutsch85},
clustering appears to be associated with a weak inertia regime, in which
the separation of the trajectories of the particles and the fluid elements
they cross in their motion, is small on the scale of the trajectory evolution.
This corresponds to a situation in which the relaxation time of the particle velocity
is shorter than the characteristic time of the random field fluctuations.
In the absence of molecular 
diffusion, the resulting clusters are of singular nature, concentrated on 
a set of zero measure \cite{deutsch85,bec03}.
 
An approach that has been fruitful in the 
study of passive tracer transport, is that of considering uncorrelated
in time random velocity fields, the so called Kraichnan model \cite{kraichnan94}.
The role of characteristic time of the random field is played in this case by
the diffusion time of a tracer (or, depending on the problem, pairs of tracers) 
across a correlation length of the field. Weak and strong inertia will then 
refer to fast or slow relaxation with respect to this characteristic
time.
The Kraichnan model approach, in the case of inertial particle transport, has  
allowed derivation of analytical expressions for the particle concentration 
correlations, both in the weak \cite{bec08,wilkinson07} and large \cite{olla08}
inertia limits. 
As recognized in \cite{bec08}, the
weak inertia limit, in a Kraichnan model approach, corresponds to a regime
of adiabatic variation for the particle separation. In this framework, the particle
concentration dynamics can be cast in the form of a problem of fast variable
elimination, in which the fast variables are the particle velocities.

A mechanism that has been proposed for cluster formation in turbulence,
is the centrifugal force induced, preferential concentration 
of heavy (light) particles in
the strain (vorticity) regions of the fluid \cite{maxey87}.
This preferential concentration effect was later
confirmed in numerical simulations \cite{wang93,hogan01}. 
Now, clustering turns out to occur also in 1D and in the Kraichnan
model just discussed, in situations therefore, where it is not clear what meaning should 
be given to preferential concentration. In particular, the concepts of
strain and vorticity do not exist in 1D. This casts some doubts on 
whether preferential concentration (or some generalized version of it)
is an essential ingredient for inertial particle clustering in 
random flows.

In general, preferential concentration could be defined as the fact
that averages of fluid (random flow) quantities, obtained from
sampling along inertial particles trajectories, do not coincide
with what would be obtained from spatial (or temporal) averages.
In other words, it could be interpreted as a non-ergodicity property of the process of
random field sampling by the particle flow.

Given the situation, a first question that would be interesting to ask,
is whether the clustering of inertial particles in random flows is 
always the result of the non ergodic sampling, by the particles, of
some relevant field quantity. In the presence of finite correlation 
time and incompressibility, physical considerations allowed to 
identify from the start, strain and vorticity as the relevant quantities. 
In the general case, such an operation may be not as easy, and an
interesting question is, therefore, whether the correct relevant quantities 
could be identified directly from the equations of motion.

This paper will try to answer these two questions.
The analysis will show that the answer is
part of the procedure of fast variable elimination required to 
determine the dynamics of the particle concentration field.
It is actually the way in which the fast variable elimination
procedure handles the presence of memory in the original process.


This paper is organized as follows. In Sec. \ref{sec2}, the equations for the
transport of a pair of inertial particle in a random field are derived. In Sec. \ref{sec3},
evolution equations for the particle separation distribution are derived and applied
to the determination of the clustering strength. In Sec. \ref{sec4} the issue of
preferential concentration and its contribution to clustering is discussed. Section
\ref{sec5} is devoted to conclusions. Technical details are left in the Appendices.

\section{Model equations}
\label{sec2}
The motion of an inertial particle in a random velocity field $\u(\x,t)$ can be modelled by the 
Stokes equation 
\beq 
\dot\v=\tau_S^{-1}[\u(\x,t)-\v],\qquad \dot\x=\v.
\label{Stokes}
\eeq
In fluid mechanics, this model would describe the dynamics of a particle that is 
sufficiently small, of sufficiently high density \cite{maxey83},
and in flow conditions in which the effect of gravity can be neglected. The relaxation 
time $\tau_S$, called the Stokes time, in the case of a spherical particle, is given by
$\tau_S=2/9\,a^2\lambda/\nu_0$, with $a$ is the particle radius, $\lambda$ the ratio
of the particle and fluid densities and $\nu_0$ the fluid kinematic viscosity.

Consider a smooth, $D$-dimensional Gaussian velocity field $\u(\x,t)$, with 
stationary and spatially uniform and isotropic statistics. 
For an incompressible random field,
the structure function $\langle\hat u_\alpha(\r,t)\hat u_\beta(\r,0)\rangle$,
$\hat\u(\r,t)\equiv \u(\x+\r,t)-\u(\x,t)$, can be written in the form
$\langle\hat u_\alpha(\r,t)\hat u_\beta(\r,0)\rangle=F(t)\hat g_{\alpha\beta}(\r)$,
where, for $r\ll\rv$:
\beq
\hat g_{\alpha\beta}(\r)=\frac{\sigma_u^2\tau_E}{\rv^2}
\Big[(D+1)r^2\delta_{\alpha\beta}-2r_\alpha r_\beta\Big],
\label{hatg}
\eeq
and $\int_0^\infty|F(t)|\d t=1$, with $\sigma_u$, $\rv$ and $\tau_E$, 
respectively, the characteristic velocity, length and time scales of the field. 
Following \cite{duncan05}, the following dimensionless parameters
are introduced:
\beq
S=\tau_S/\tau_E,
\qquad
K=\sigma_u\tau_E/r_\v,
\label{Stokes-Kubo}
\eeq
called respectively Stokes and Kubo numbers.
Units are chosen such that $\sigma_u=\tau_S=1$; in this way:
\beq
\tau_E=S^{-1},
\qquad
r_\v=(KS)^{-1}.
\label{tauE-rv}
\eeq
The Kubo number describes the intrinsic long- or short-correlated in time nature of
the field, with $K\to 0$ ($S\to\infty$) corresponding to an uncorrelated regime: 
$F(t)\to 2\delta(t)$; $K\to\infty$ ($S\to 0$) corresponds in turn
to frozen field conditions. For real turbulence, $K\sim 1$ and $S$
parameterizes the strength of inertia.

The $K\to 0$ regime, corresponding to a Kraichnan
model for inertial particles, allows to neglect the particle displacement in a correlation
time $\tau_E$. 
Clustering is a phenomenon associated with singular behavior, at small
values of the argument, of the particle separation PDF (probability density function) $\rho(\r,t)$.
A Kraichnan model approach allows to describe the two-particle dynamics relevant for
clustering, in terms of a single equation:
\beq 
\dot\bnu=\hat\u(\r,t)-\bnu,\qquad \dot\r=\bnu,
\label{Stokes2}
\eeq
as the evolution of $\hat\u(\r,t)$ depends only on $\r$, and not separately on the coordinates
of the two particles.
[For finite $\tau_E$, this would not be true, as $\hat\u(r(t),t)$ would depend, on scale $\tau_E$,
on the separate evolution of the two particle coordinates, described by Eq. (\ref{Stokes})].
In a Kraichnan model approach, the separation PDF $\rho(\r,t)$ will obey 
the Fokker-Planck equation associated with Eq. (\ref{Stokes2}) (more precisely, its
restriction to the variable $\r$).

As recognized in \cite{duncan05}, the two-particle dynamics becomes dependent, in an uncorrelated in time regime, on the single parameter
\beq
\epsilon=\frac{\sigma_u^2\tau_E}{\rv^2}=K^2S,
\label{epsilon}
\eeq
that is the amplitude factor in front of Eq. (\ref{hatg}). This parameter 
plays the role of generalized Stokes number 
for a Kraichnan model. Writing  
$\epsilon=\tau_S/\tilde\tau_E$, one notices in fact that 
$\tilde\tau_E=\rv^2/(\sigma_u^2\tau_E)$ is the time for a 
tracer [a point moving with velocity $\u(\x(t),t)$] to diffuse across $\rv$, that
plays the role of characteristic time scale for the uncorrelated in time random flow.

Small $\epsilon$  corresponds to a weak inertia regime, and Eqs. 
(\ref{hatg},\ref{Stokes2},\ref{epsilon})
provide, for $\tau_E\to 0$, 
a Kraichnan model for the clustering of weakly inertial particles
(see \cite{duncan05,wilkinson07,bec08} and references therein).
To see that small $\epsilon$ corresponds to weak inertia, it suffices to 
verify that $r$ changes little on a time $\tau_S$ (the correlation 
time for $\bnu$): $\Delta r(\tau_S)\ll r$.
This is verified a-posteriori solving Eqs. (\ref{Stokes2},\ref{hatg})
for fixed $\r$, $r\ll\rv$. This gives $\langle\nu^2|r\rangle\sim\epsilon r^2$,
that coincides, in the dimensionless units of Eq. (\ref{tauE-rv}), with
the square displacement in a time $\tau_S$. 
Thus, $\Delta r(\tau_S)/r\sim\epsilon^{-1/2}\ll 1$, as expected.

This is an adiabatic regime for $\r$, that has allowed the authors in 
\cite{bec08}
to use a fast variable elimination technique to derive a version of the Fokker-Planck 
equation associated with Eq. (\ref{Stokes2}), restricted to $\r$. In the following sections, 
a similar approach will be utilized to establish the connection between clustering
and preferential concentration effects.

\section{Clustering}
\label{sec3}
The fast variable elimination procedure in a stochastic problem, like the one described
by Eq. (\ref{Stokes2}) in the $\epsilon\ll 1$ regime, can be carried on substantially
in two ways \cite{lindenberg}: working at the Fokker-Planck equation level, by means
of so called projection operator techniques \cite{zwanzig61,grigolini85}, or averaging away
the fast variables already at the level of the stochastic differential equation. 
The procedure followed in \cite{bec08} was of the projection operator type, along
the lines of the approach derived in \cite{majda01}, in the context of stochastic
climate modelling. This approach is not appropriate here: 
dealing with preferential concentration effects will require evaluation of
conditional averages such as $\langle f_n[\hat\u]|\r(t)$=$\bar\r\rangle$, with $f_n[\hat\u]$
products of the fields and their derivatives, about which, the 
Fokker-Planck equation associated with Eq. (\ref{Stokes2}) provides no information.

The approach that is going to be followed here, is the one described in \cite{sancho82,hanggi85},
based on the use of the so called stochastic Liouville equation, and of functional derivation
techniques \cite{furutsu63,novikov65}. A similar approach was followed in \cite{olla08}, 
to analyze the large inertia limit of particle transport.

The evolution equation for the PDF $\rho(\bar\r,t)$, associated with Eq. (\ref{Stokes2}), 
can be written in the form
$\partial_t\rho(\bar\r,t)+\bar\partial_\alpha J_\alpha(\bar\r,t)=0$,
$\bar\partial_\alpha\equiv\partial/\partial\bar r_\alpha$,
with the probability current $J_\alpha$ given by
\beq
J_\alpha(\bar\r,t)=
\int_{-\infty}^t\d\tau\,\ex^{\tau-t}
\langle\hat u_\alpha(\r(\tau),\tau)\delta(\r(t)-\bar\r)\rangle.
\label{current}
\eeq
This evolution equation is basically
the average over all the realizations of $\hat\u$,
of the Liouville equation in the configuration space of Eq. (\ref{Stokes2}):
$\partial_t\tilde\rho(\bar\r,t)+\bar\partial_\alpha[\nu_\alpha(t)\tilde\rho(\bar\r,t)]=0$,
$\tilde\rho(\bar\r,t)=\delta(\r(t)-\bar\r)$. 
The integral
$\int_{-\infty}^t\d\tau\ex^{\tau-t}\hat\u(\r(\tau),\tau)$ in Eq. (\ref{current}), 
in fact, is just the solution for $\bnu(t)$ 
of Eq. (\ref{Stokes2}), and ${\bf J}\equiv
\langle\bnu(t)|\r(t)$=$\bar\r\rangle\rho(\bar\r,t)$, 
with $\rho(\bar\r,t)=\langle\tilde\rho(\bar\r,t)\rangle$.


The correlation between the Dirac delta and the random field in Eq. (\ref{current}) 
is calculated using the
functional integration by part formula \cite{furutsu63,novikov65} (see also \cite{frisch}).
Indicating
by $\delta/\delta\hat\u(\x,t)$ the operation of functional differentiation, this
corresponds to making in Eq. (\ref{current}) the substitution
\beq
\hat u_\alpha(\z,t)\longrightarrow
2\int\d^D r'\hat g_{\alpha\beta}(\z,\z')
\frac{\delta}{\delta\hat u_\beta(\z',t)},
\label{bypart}
\eeq
where $2\hat g_{\alpha\beta}(\z,\z')\delta(t-t')
=\langle\hat u_\alpha(\z,t)\hat u_\beta(\z',t')\rangle$
and of course $\hat g_{\alpha\beta}(\z,\z)\equiv\hat g_{\alpha\beta}(\z)$ \cite{nota1}.
In order to use Eq. (\ref{bypart}), however, it is first necessary to  
write 
in Eq. (\ref{current}) $\hat u_\alpha(\r(\tau),\tau)=\int\d^D z\hat u_\alpha(\z,\tau)
\delta(\r(\tau)-\z)$; the functional derivative 
$\delta/\delta\hat u_\beta(\z',\tau)$, will then act on a product 
$\delta(\r(\tau)-\z)\delta(\r(t)-\bar\r)$. This requires determination of 
expressions like
\beq
R_{\gamma\beta}(t;\z',\tau)=\frac{\delta r_\gamma(t)}{\delta\hat u_\beta(\z',\tau)}.
\label{response}
\eeq
Writing $\delta/\delta\hat u_\beta(\z',\tau)\delta(\r(0)-\bar\r)
=-R_{\gamma\beta}(t;\z',\tau)\bar\partial_\gamma\delta(\r(t)$ $-\bar\r)$, it is clear
that the role of the response function $R_{\gamma\beta}(t;\z',\tau)$ in Eq. (\ref{current})
is precisely to account 
for the correlation between the random field and the condition on the separation
at time $t$.

Following the approach in \cite{sancho82}, outlined in Appendix A, the response function
can be calculated as an expansion in powers of $\hat\u$ [i.e., from Eqs. 
(\ref{hatg}) and (\ref{epsilon}), basically in powers of $\epsilon^{1/2}$]:
$R_{\gamma\beta}(t;\z,\tau)=
\delta(\r(\tau)-\z)[\hat R_{\gamma\beta}^\smalze+\hat R_{\gamma\beta}^\smalun+\ldots]$. The
first two terms in the expansion read [see Eqs. (\ref{A3}-\ref{A4})]: 
\begin{eqnarray}
\hat R_{\gamma\beta}^\smalze&=&\psi(\tau-t)\delta_{\gamma\beta},
\quad
\psi(t)=\theta(-t)[1-\ex^t],
\label{response0}
\\
\hat R_{\gamma\beta}^\smalun&=&
\int_\tau^t\d\tau'\psi(\tau-\tau')
\psi(\tau'-t)
\frac{\partial\hat u_\gamma(\r(\tau'),\tau')}{\partial r_\beta(\tau')}
\label{response1}
\end{eqnarray}
and $\theta(t)$ is the Heaviside step function
[$\theta(t)=1$ for $t>0$, $\theta(t)=0$ otherwise]. 

From Eqs. (\ref{response0}-\ref{response1})
and (\ref{A4}), it appears that $R_{\gamma\beta}(\tau;\z',\tau)=0$, so that
the functional derivative on $\delta(\r(\tau)-\z)$, arising in Eq. (\ref{current}) from
$\hat u_\alpha(\r(\tau),\tau)=\int\d^D z\hat u_\alpha(\z,\tau)
\delta(\r(\tau)-\z)$, 
does not contribute.
The analysis in Appendix A shows that the lowest order inertia contribution to $J_\alpha$ occurs
at $O(\epsilon^2)$, corresponding to taking into account only 
$R^\smalun_{\gamma\beta}(t;\z',\tau)$
in the expansion for
$R_{\gamma\beta}(t;\z',\tau)$.
Carrying out the necessary functional differentiations in Eq. (\ref{current}) and using 
Eqs. (\ref{response0}-\ref{response1}) gives then the result, for $t=0$:
\begin{eqnarray}
J_\alpha&\simeq&-2\bar\partial_\gamma\int_{-\infty}^0\d\tau\ex^\tau
\Big\langle\hat g_{\alpha\beta}(\r(\tau))
\delta(\r(0)-\bar\r)
\Big[\delta_{\gamma\beta}\psi(\tau)
\nonumber
\\
&+&\int_\tau^0\d\tau'\psi(\tau-\tau')\psi(\tau')
\frac{\partial\hat u_\gamma(\r(\tau'),\tau')}{\partial r_\beta(\tau')}\Big]\Big\rangle.
\label{step2}
\end{eqnarray}
The lowest order contribution to the current is, exploiting incompressibility,
$\bar\partial_\gamma\hat g_{\alpha\gamma}(\bar\r)=0$:
\beq
J^\smaldu_\alpha=-g_{\alpha\gamma}(\bar\r)\bar\partial_\gamma\rho(\bar\r,t),
\label{order2}
\eeq
corresponding,
in Eq. (\ref{step2}), 
to put $\hat g_{\alpha\beta}(\r(\tau))\simeq\hat g_{\alpha\beta}(\r(0))$ 
and to neglect the term $\propto \partial_\beta\hat u_\gamma$ in the
integral. Equation (\ref{order2}) describes pure tracer transport.

The next order is $J^\smalqu$, that receives two contributions: the one 
$\propto  \partial_\beta\hat u_\gamma$, originating from $\hat R^\smalun$, and one
$\propto \hat g_{\alpha\beta}(\r(\tau))-\hat g_{\alpha\beta}(\r(0))$.
The first contribution is calculated applying Eq. (\ref{bypart}) to the factor 
$\partial\hat u_\gamma(\r(\tau'),\tau')/\partial r_\beta(\tau')$ in (\ref{step2}).
The second contribution is calculated writing 
$\hat g_{\alpha\beta}(\r(\tau))-\hat g_{\alpha\beta}(\r(0))$ as a function of 
$\Delta\r(\tau)=\r(\tau)-\r(0)$ and using the
expression, which descends from Eq. (\ref{Stokes2}):
\beq
\Delta\r(\tau)=\int_{-\infty}^0\d\tau'\Phi(\tau',\tau)\hat\u(\r(\tau'),\tau'),
\label{Deltar}
\eeq
where 
\beq
\Phi(\tau',\tau)=\psi(\tau'-\tau)-\psi(\tau'). 
\label{Phi}
\eeq
Evaluating to $O(\epsilon^2)$ the two contributions and exploiting incompressibility, 
leads to the result (see Appendix B): 
\beq
J^\smalqu_\alpha=-\frac{1}{2}\rho(\bar\r)
\bar\partial_\beta\bar\partial_\eta\hat g_{\alpha\gamma}(\bar\r)
\bar\partial_\gamma\hat g_{\beta\eta}(\bar\r)
\label{order4}
\eeq
plus other terms containing spatial derivatives of $\rho$.
The stationary solution $\rho(\bar\r)$ is obtained requiring that the current
be divergence free: $\partial_\alpha J_\alpha=0$. The lowest order solution,
from Eq. (\ref{order2}), is spatially uniform, and the derivatives acting
on $\rho$ do not contribute in Eq. (\ref{order4}), at the order considered. 
The stationary solution 
will obey, therefore, the equation
\beq
\hat g_{\alpha\gamma}(\bar\r)\bar\partial_\alpha\bar\partial_\gamma\rho(\bar\r)+
\frac{\rho(\bar\r)}{2}\bar\partial_\beta\bar\partial_\eta\hat g_{\alpha\gamma}(\bar\r)
\bar\partial_\alpha\bar\partial_\gamma\hat g_{\beta\eta}(\bar\r)=0,
\label{stationary}
\eeq
that is
in agreement with \cite{bec08}, once a wrong sign in their Eq. (40) is corrected 
\cite{turitsyn09}. As discussed in \cite{wilkinson07,bec08}, 
the solution of Eq. (\ref{stationary}), with the expression for $\hat g_{\alpha\beta}(\r)$ 
provided in Eq. (\ref{hatg}), is a power law  at $r\to 0$: $\rho(\r)\propto r^{-c\epsilon}$,
$c=2(D+1)(D+2)$, corresponding to the correlation dimension for the particle distribution:
$D_2=D-2(D+1)(D+2)\epsilon$.

\section{Preferential concentration}
\label{sec4}
In the previous section, clustering was obtained by solving an evolution equation for
the separation PDF $\rho(\bar\r,t)$: $\partial_t\rho+\bar\partial_\alpha J_\alpha=0$,
in which the probability current divergence $\bar\partial_\alpha J_\alpha$ was expressed, 
through Eq. (\ref{current}), in terms of a conditional average
$\bar\partial_\alpha\langle\hat u_\alpha(\r(\tau),\tau)\delta(\r(0)-\bar\r)\rangle$.
Since, in the absence of conditioning, this average would be zero, it is clear
that non-ergodic sampling of the random field along the particle trajectories
is a necessary condition for a non-zero current, and hence for clustering.

In order to clearly identify the quantities undergoing preferential 
concentration, it is necessary to pass the divergence operator in 
$\bar\partial_\alpha\langle\hat u_\alpha(\r(\tau),\tau)\delta(\r(0)-\bar\r)\rangle$,
inside the average, so that the expression can be written in the form
$\langle f[\hat\u]\delta(\r(0)-\bar\r)\rangle$. From $f[\hat\u]$, one can then extract
the contributions from specific monomials $f_n[\hat\u]$ in the fields and their
derivatives.

In the present situation, a good strategy to identify these monomials
is to expand the field $\hat u_\alpha(\r(\tau),\tau)$ around the final
point $\r(0)$:
\begin{eqnarray}
\hat\u(\r(\tau),\tau)=\{1+\Delta r^\smalun_\beta(\tau)\partial_\beta
+\Delta r^\smaldu(\tau)\partial_\beta
\nonumber
\\
+\frac{1}{2}\Delta r^\smalun_\beta(\tau)\Delta r^\smalun_\gamma(\tau)
\partial_\beta\partial_\gamma+\ldots\}
\hat\u(\r(0),\tau),
\label{expansion}
\end{eqnarray}
where
\begin{eqnarray}
\Delta\r^\smalun(\tau)&=&\hat\Phi\hat\u(\r(0),\tau),
\label{Deltar1}
\\
\Delta\r^\smaldu(\tau)&=&\hat\Phi
\Delta r^\smalun_\beta\partial_\beta\hat\u(\r(0),\tau),
\label{Deltar2}
\end{eqnarray}
and so on to higher orders, 
with $\hat\Phi g(\tau)\equiv\int_{-\infty}^0\d\tau'\Phi(\tau',\tau)$ $\times g(\tau')$ and
$\Phi(\tau,\tau')$ given in Eq. (\ref{Phi}). In the weak inertia regime considered, 
$\epsilon\ll 1$, this is appropriate in the 
interval $-\tau\le 1$, in which the time integral in Eq. (\ref{current}) is
concentrated.

Substituting  Eqs. (\ref{expansion}-\ref{Deltar2}) into  (\ref{current}),
the current divergence can be written as a sum of averages, involving increasing powers
of $\hat\u$. 
The first contributions are:
\begin{eqnarray}
f_1[\hat\u]&=&\bar\partial_\alpha\hat u_\alpha(\bar\r,\tau),
\label{f1}
\\
f_2[\hat\u]&=&\bar\partial_\alpha\bar\partial_\beta[\hat u_\alpha(\bar\r,\tau)
\hat u_\beta(\bar\r,\tau')],
\label{f2}
\\
f_3[\hat\u]&=&\bar\partial_\alpha\bar\partial_\beta\bar\partial_\gamma[
\hat u_\alpha(\bar\r,\tau)\hat u_\beta(\bar\r,\tau')\hat u_\gamma(\bar\r,\tau'')].
\label{f3}
\end{eqnarray}
Other contributions involve gradients of the Dirac delta in Eq. (\ref{current}),
that would lead in the end to gradients of $\rho(\bar\r,t)$. The terms that
would lead to clustering even starting from a spatially homogeneous particle
distribution, however, are those in Eqs. (\ref{f1}-\ref{f3}).

The contribution from Eq. (\ref{f1}) is evaluated following the same procedure
leading from Eq. (\ref{current}) to (\ref{step2}). Stopping to lowest order in 
$\epsilon$ and neglecting terms involving gradients of $\rho$,
leads to the result (see Appendix C):
\beq
\langle \bar\partial_\alpha\hat u_\alpha(\bar\r,\tau)|\r(0)=\bar\r\rangle
=-\frac{1}{2}\psi(\tau)\bar\partial_\alpha\bar\partial_\gamma\hat g_{\alpha\gamma}(\bar\r),
\label{<f1>}
\eeq
that will vanish if $\u$ is incompressible.
This means that, in the case of a compressible random field, particularly in $D=1$,
clustering will be the result of preferential concentration in regions of negative divergence 
of $\u$. More precisely, particle pairs that at time $t=0$ have separation $\bar\r$, in 
the past were more likely to be in regions of negative $\nabla\cdot\u$.

It is to be noticed that Eq. (\ref{expansion}) 
does not correspond to an expansion of $J_\alpha$ in powers of $\epsilon^{1/2}$.
The fields $\hat\u$ in Eq. (\ref{expansion}), upon substitution in Eq. (\ref{current}),
contribute $O(\epsilon^{1/2})$ in contraction
with another field $\hat\u$, but contribute $O(\epsilon)$ in contraction
with $\delta(\r(0)-\bar\r)$. A consequence of this is that Eq. (\ref{<f1>}) 
does not provide the whole $O(\epsilon)$ part of $\bar\partial_\alpha J_\alpha$.
In the compressible case, another $O(\epsilon)$ contribution to $\bar\partial_\alpha J_\alpha$
is provided for instance by
$\bar\partial_\alpha\bar\partial_\beta
\hat g_{\alpha\beta}(\bar\r)$, that comes from the part of
$\langle\bar\partial_\alpha\bar\partial_\beta[\hat u_\alpha(\bar\r,\tau)
\hat u_\beta(\bar\r,\tau')]|\r(0)$=$\bar\r\rangle$ that is uncorrelated with 
$\delta(\r(0)-\bar\r)$.

In the incompressible case, the contribution from $f_1$ 
to the current divergence vanishes, and it is necessary to consider $f_n$, 
with $n>1$; in particular, the term in Eq. (\ref{f2}).
Now, the strain 
and the vorticity variance of a velocity field $\u$ are defined from
\begin{align}
|{\bf S}|^2&=(1/2)\partial_\alpha u_\beta\,(\partial_\alpha u_\beta+
\partial_\beta u_\alpha)
\nonumber
\\
|\bomega|^2&=\partial_\alpha u_\beta\,(\partial_\alpha u_\beta-
\partial_\beta u_\alpha),
\label{strain-vorticity}
\end{align}
and $\partial_\alpha u_\beta\partial_\beta u_\alpha=|{\bf S}|^2-|\bomega|^2/2$,
whose average can be shown to be zero if $\u$ is incompressible. 
Thus, $f_2[\hat\u]$ is connected with the difference $|{\bf S}|^2-|\bomega|^2/2$
for $\u$.

Since, in the case of an incompressible field:
$\langle|{\bf S}|^2\rangle=\langle|\bomega|^2/2\rangle$,
$\langle f_2[\hat\u]|\r(0)$=$\bar\r\rangle$ 
will receive its first non-zero contribution by contraction of the two fields 
$\hat u_\alpha$ and $\hat u_\beta$ in $f_2$ with $\delta(\r(0)-\bar\r)$
[see Eq. (\ref{f2})]. 
The calculation carried on in Appendix C gives the result, for $\tau>\tau'$:
\begin{eqnarray}
\langle\bar\partial_\alpha\hat u_\beta(\bar\r,\tau')
\bar\partial_\beta\hat u_\alpha(\bar\r,\tau)|\r(0)=\bar\r\rangle
\nonumber
=
\psi(\tau)
\\
\times
[3\psi(\tau')-\psi(\tau'-\tau)]
\bar\partial_\alpha\bar\partial_\beta\hat g_{\gamma\eta}(\bar\r)
\bar\partial_\gamma\bar\partial_\eta\hat g_{\alpha\beta}(\bar\r),
\label{<f2>}
\end{eqnarray}
plus terms involving gradients of $\rho(\bar\r,t)$.
The quantity to RHS (right hand side) of Eq. (\ref{<f2>}) is positive for $\tau'<\tau<0$.
Thus, preferential concentration in regions of high strain, in the wider sense
given here to the term, is a property that is mantained in an incompressible 
Kraichnan model for inertial particle transport.

Again, Eq. (\ref{<f2>}) does not account for the whole of the current divergence
to $O(\epsilon^2)$ in the incompressible case. One has to consider also the
skewness-like term in Eq. (\ref{f3}). This time, it is necessary to consider
the contraction of only one of the fields with $\delta(\r(0)-\bar\r)$.
A calculation completely analogous to that of Eq. (\ref{<f1>}) gives then the result:
\begin{eqnarray}
&&\langle
\bar\partial_\alpha\bar\partial_\beta\bar\partial_\gamma[
\hat u_\alpha(\bar\r,\tau)\hat u_\beta(\bar\r,\tau')\hat u_\gamma(\bar\r,\tau'')]
|\r(0)=\bar\r\rangle
\nonumber
\\
&&=-\bar\partial_\alpha\bar\partial_\beta\hat g_{\gamma\eta}(\bar\r)
\bar\partial_\gamma\bar\partial_\eta\hat g_{\alpha\beta}(\bar\r)[\psi(\tau)\delta(\tau'-\tau'')
\nonumber
\\
&&+{\rm permutations}],
\label{<f3>}
\end{eqnarray}
plus, again,  terms involving gradients of $\rho(\bar\r,t)$.
Thus, clustering receives, in the incompressible case, contribution from 
preferential concentration in regions of simultaneously high strain and negative  
skewness $\bar\partial_\alpha\bar\partial_\beta\bar\partial_\gamma[
\hat u_\alpha(\bar\r,\tau)\hat u_\beta(\bar\r,\tau')\hat u_\gamma(\bar\r,\tau'')]$.

The terms in Eqs. (\ref{<f2>}-\ref{<f3>}) 
are all the preferential concentration contributions that arise, at $O(\epsilon^2)$,
in the incompressible case. 
A term $f_4$ of fourth order in $\hat\u$, 
not indicated in Eq. (\ref{expansion}),
may contribute 
to $\langle f_4[\hat\u]|\r(0)$=$\bar\r\rangle$; 
at $O(\epsilon^2)$, this occurs
through the unconditional average
$\langle f_4[\hat\u]\rangle$. 
However, Gaussianity of $\u$ reduces this contribution 
to products $\langle f_2[\hat\u]\rangle \langle f_2[\hat\u]\rangle$,
which vanish in the case of an incompressible random field. 

It is to be noticed that the RHS of Eqs. (\ref{<f1>}-\ref{f3}) are independent of the argument
$\bar\r$. This means that the averages in those equations are conditioned to the presence
of a pair of particles at unspecified separation. 
In other words, the two-particle conditional averages in 
Eqs. (\ref{<f1>}-\ref{<f3>}), are equivalent
to averages conditioned to a single particle in $\x+\bar\r$.

\section{Conclusion}
\label{sec5}
The analysis in this paper shows that some generalized form of 
preferential concentration continues to be an ingredient for 
clustering, also in situations where its presence is  not expected,
such as transport by uncorrelated in time, or 1D random velocity fields. 
The original statement, that
the random field at the particle location has on the average some
property (e.g. higher strain), is replaced, in the uncorrelated case, by 
one on properties of the random field, still at that
location, but at times previous to the arrival of the particle.
The current field configuration, instead, is uncorrelated with 
the current particle positions. In other words, if the field
is uncorrelated in time, there will be no preferential concentration, 
in the standard meaning of the term,
rather, a generalized version 
involving field and particle 
configurations at different times.


The weak inertia considered, has allowed to identify the
quantities involved in this form of generalized preferential concentration, 
through 
an expansion in the effective Stokes number $\epsilon$ [see Eqs. 
(\ref{Stokes-Kubo},\ref{tauE-rv},\ref{epsilon})].
In the two regimes of compressible and incompressible flow,
the relevant field properties are related to the field
divergence at the particle position, in the first case, 
and to the difference between the square strain and vorticity
in the second. An additional relevant quantity in the
incompressible case, not previously considered in the
literature (to the author's knowledge), is the skewness 
in Eq. (\ref{<f3>}).

The standard interpretation of preferential concentration is recovered
considering a local time average of the flow:
$\bar\u(\bar\r,t)=\Delta t^{-1}\int_{-\Delta t/2}^{\Delta t/2}\hat\u(\bar\r,t+\tau)\d\tau$.
Basically, this gives to the random field a finite correlation time $\Delta t$.
From Eqs. (\ref{f1}-\ref{f3}), it is then possible to define
time averaged quantities
$\bar\partial_\alpha\bar u_\alpha(\bar\r,\tau)$,
$\bar\partial_\alpha\bar\partial_\beta[\bar u_\alpha(\bar\r,\tau)
\bar u_\beta(\bar\r,\tau)]$, and
$\bar\partial_\alpha\bar\partial_\beta\bar\partial_\gamma[
\bar u_\alpha(\bar\r,\tau)\bar u_\beta(\bar\r,\tau)\bar u_\gamma(\bar\r,\tau)]$, 
corresponding to divergence, difference between strain and vorticity squared
[see Eq. (\ref{strain-vorticity})] and skewness of the field $\bar\u$. From Eqs. 
(\ref{<f1>}-\ref{<f3>}), it is easy to see that inertial particles,
with respect to these quantities,
undergo preferential concentration in the usual sense of the word.

The analogy between particle transport by a Kraichnan model 
and by a random velocity field with finite correlation time,
with the parameter $\epsilon$ in the first case set equal to $S$ in the
second,
allows predictions of the preferential concentration 
strength in the finite correlation time case.
Indicating by $\langle.\rangle_p$,
average at a particle position:
$(\rv/\sigma_u)\langle\nabla\cdot\u\rangle_p=O(S^{1/2})$, in the
compressible case, and 
$(\rv/\sigma_u)^2\langle (|{\bf S}|^2-|\bomega|^2/2)\rangle_p=O(S)$,
$(\rv/\sigma_u)^3\langle\partial_\alpha\partial_\beta\partial_\gamma[
u_\alpha u_\beta u_\gamma]\rangle_p=O(S^{1/2})$
in the incompresible one. Preferential concentration should 
occur in regions of 
negative field divergence in the compressible case,
high strain and negative skewness in the incompressible one.

The techniques utilized in this paper 
are not specific to inertial particles, and
are in fact rather general.
For instance, conclusions analogous to the 
ones on inertial particles,
could be drawn,
in the case of fluid tracers
in compressible flows, 
both regarding clustering
and preferential concentration.
(In fact, it would be interesting to understand how much of 
particle clustering in compressible flows is associated with inertia, and 
how much is due to accumulation on fluid shocks).

\begin{acknowledgements} The author wishes to thank K. Turitsyn for interesting and
helpful discussion.
\end{acknowledgements}

\appendix
\setcounter{section}{1}
\setcounter{equation}{0}
\section*{Appendix A. Response function determination}
Following \cite{sancho82}, an equation for the response function 
$R_{\gamma\beta}(t;\z,\tau)=\delta r_\gamma(t)/\delta\hat u_\beta(\z,\tau)$ 
can be obtained writing Eq. (\ref{Stokes2}) in the form
$$
\ddot r_\gamma(t)+\dot r_\gamma(t)=\hat u_\gamma(\r(t),t)
$$
and taking the functional derivative with respect to $\hat u_\beta(\z,\tau)$.
The result is, for $t>\tau$:
\beq
\ddot R_{\gamma\beta}(t;\z,\tau)+\dot  R_{\gamma\beta}(t;\z,\tau)=
\frac{\partial\hat u_\gamma(\r(t),t)}{\partial r_\eta(t)}R_{\eta\beta}(t;\z,\tau).
\label{A1}
\eeq
From Eq. (\ref{Stokes2}), one can write
$\r(t)={\rm const.}+\int_{-\infty}^t\d\tau\psi(\tau-t)\hat\u(\r(\tau),\tau)$
and $\bnu(t)=\int_{-\infty}^t\d\tau\ex^{\tau-t}\hat\u(\r(\tau),\tau)$,
with $\psi(t)$ given in Eq. (\ref{response0}). 
These expressions lead to the initial conditions
\beq
R_{\gamma\beta}(\tau;\z,\tau)=0;
\quad\ 
\dot R_{\gamma\beta}(\tau;\z,\tau)=\delta_{\gamma\beta}\delta(\r(\tau)-\z).
\label{A2}
\eeq
Equation (\ref{A1}) with the initial conditions in Eq. (\ref{A2}) can 
be solved perturbatively in $\hat\u$ (i.e. basically in $\epsilon^{1/2}$):
$R_{\gamma\beta}(t;\z,\tau)=\delta(\r(\tau)-\z)[\hat R_{\gamma\beta}^\smalze+
\hat R_{\gamma\beta}^\smalun+\ldots]$, with the result
\begin{eqnarray}
\hat
R_{\gamma\beta}^\smalze&=&\psi(\tau-t)\delta_{\gamma\beta},
\label{A3}
\\
\hat R_{\gamma\beta}^\smalnp1&=&
\int_\tau^t\d\tau'\psi(\tau'-t)
\frac{\partial\hat u_\gamma(\r(\tau'),\tau')}{\partial r_\eta(\tau')}
\hat R_{\eta\beta}^\smaln.
\label{A4}
\end{eqnarray}
In particular, the second order term reads:
\begin{eqnarray}
\hat R_{\gamma\beta}^\smaldu&=&
\int_\tau^t\d\tau'\int_\tau^{\tau'}\d\tau''\psi(\tau-\tau'')\psi(\tau''-\tau')\psi(\tau'-t)
\nonumber
\\
&\times&
\frac{\partial\hat u_\gamma(\r(\tau'),\tau')}{\partial r_\phi(\tau')}
\frac{\partial\hat u_\phi(\r(\tau''),\tau'')}{\partial r_\beta(\tau'')}.
\label{A5}
\end{eqnarray}
It is possible to see that, to $O(\epsilon^2)$, this term does not contribute 
to the current $J_\alpha$. In fact, including $\hat R_{\gamma\beta}^\smaldu$ in Eq. 
(\ref{step2}), the contraction $\partial_\phi\hat u_\gamma\partial_\beta\hat u_\phi\propto
\epsilon\delta(\tau'-\tau'')$ would be killed by the factor $\psi(\tau''-\tau')$ from 
Eq. (\ref{A5}). On the other hand, the contractions of the fields $\partial_\phi\hat u_\gamma$
and $\partial_\beta\hat u_\phi$, with the other factors in Eq. (\ref{step2}), would produce
additional factors $\epsilon^{1/2}$, beyond those from the fields themselves, and the
final result would be of higher order in $\epsilon$.

\setcounter{section}{2}
\setcounter{equation}{0}
\section*{Appendix B. Contributions to the probability current}
The contribution from $\hat R^\smalun$ in Eq. (\ref{step2}) is evaluated using the
functional integration by part formula Eq. (\ref{bypart}), and keeping only $\hat R^\smalze$
in the resulting expansion for the response function. Indicating by $J_\alpha^R$ 
this contribution:
\begin{eqnarray}
J^R_\alpha&=&4
\int_{-\infty}^0\d\tau\int_\tau^0\d\tau'
\int\d^Dz\int\d^Dz'\ex^\tau\psi(\tau-\tau')
\nonumber
\\
&\times&\psi^2(\tau')
\partial_{z_\beta}\hat g_{\gamma\eta}(\z,\z')
\bar\partial_\gamma\bar\partial_\eta
\langle\hat g_{\alpha\beta}(\r(\tau))
\nonumber
\\
&\times&\delta(\r(\tau')-\z)\delta(\r(\tau')-\z')\delta(\r(0)-\bar\r)
\rangle.
\label{B1}
\end{eqnarray}
Notice that, due to the condition $\tau'>\tau$, the functional derivative acted only on 
$\delta(\r(0)-\bar\r)$ and not on $\hat g_{\alpha\beta}(\r(\tau))$.
To $O(\epsilon^2)$, one sets $\r(\tau')=\r(0)$ in Eq. (\ref{B1}) and carrying out the
integrals obtains the result
\beq
J^R_\alpha=\frac{1}{3}
\rho(\bar\r)
\bar\partial_\beta\bar\partial_\eta\hat g_{\alpha\gamma}(\bar\r)
\bar\partial_\gamma\hat g_{\beta\eta}(\bar\r),
\label{B2}
\eeq
plus terms involving spatial derivatives of $\rho$.

The other contribution in Eq. (\ref{step2}) is:
\begin{eqnarray}
J^g_\alpha&=&-2\bar\partial_\gamma\int_{-\infty}^0\d\tau\ex^\tau\psi(\tau)
\nonumber
\\
&\times&
\langle[\hat g_{\alpha\gamma}(\r(\tau))-\hat g_{\alpha\beta}(\r(0))]
\delta(\r(0)-\bar\r)\rangle,
\label{B3}
\end{eqnarray}
which can be seen to contain both quadratic and linear terms in $\Delta\r$, 
with  $\Delta\r$ given in Eq. (\ref{Deltar}). Writing
\begin{eqnarray}
&\hat g_{\alpha\beta}(\r(\tau))-\hat g_{\alpha\beta}(\r(0))=
\hat G_{\alpha\beta\eta\phi}[\Delta r_\eta(\tau)\Delta r_\phi(\tau)
\nonumber
\\
&+r_\eta(0)\Delta r_\phi(\tau)+
r_\phi(0)\Delta r_\eta(\tau)],
\nonumber
\end{eqnarray}
with
$\hat G_{\alpha\beta\eta\phi}=
(1/2)\partial_\eta\partial_\phi\hat g_{\alpha\beta}(\r)$, 
substituting into Eq. (\ref{B3}) and using Eq. (\ref{Deltar}) allows to calculate
immediately the quadratic term
\beq
J^{g,2}_\alpha=
-\frac{5}{6}
\rho(\bar\r)
\bar\partial_\beta\bar\partial_\eta\hat g_{\alpha\gamma}(\bar\r)
\bar\partial_\gamma\hat g_{\beta\eta}(\bar\r),
\label{B4}
\eeq
plus terms involving, again, spatial derivatives of $\rho$; 
the linear terms lead only to contributions involving
derivatives of $\rho$. Combining Eqs. (\ref{B2}) and (\ref{B4}) leads
to Eq. (\ref{order4}).

\setcounter{section}{3}
\setcounter{equation}{0}
\section*{Appendix C. Evaluation of non-ergodic terms}
As with Eq. (\ref{current}), calculation of conditional
averages like
$\langle f_n[\hat\u]|\r(0)$=$\bar\r\rangle$, 
with $f_n[\hat\u]$ given in Eqs. (\ref{f1}-\ref{f3}),
is carried on by application of the 
functional integration by part formula Eq. (\ref{bypart}).
In the case of $f_1[\hat\u]$, one has to lowest order in $\epsilon$, 
from Eqs. (\ref{bypart}) and (\ref{response0}):
\begin{eqnarray}
&&\langle\bar\partial_\alpha\hat u_\alpha(\bar\r,\tau)\delta(\r(0)-\bar\r)\rangle
\nonumber
\\
&=&2[\bar\partial_\beta\langle\delta(\r(0)-\bar\r)\bar\partial_\alpha
\hat g_{\alpha\beta}(\bar\r,\r(\tau))\rangle
\nonumber
\\
&-&\langle\delta(\r(0)-\bar\r)\bar\partial_\alpha\bar\partial_\beta
\hat g_{\alpha\beta}(\bar\r,\r(\tau))\rangle]
\label{C1}
\end{eqnarray}
From the definition: $2\hat g_{\alpha\beta}(\r,\r')\delta(t-t')=\langle\hat u_\alpha(\r,t)
\hat u_\beta(\r',t')\rangle$, with $\hat u(\r,t)=\u(\x+\r,t)-\u(\x,t)$, it is easy to see
that $\hat g_{\alpha\beta}(\r,\r')=(1/2)[\hat g_{\alpha\beta}(\r)+
\hat g_{\alpha\beta}(\r')-\hat g_{\alpha\beta}(\r-\r')]$, and therefore
\beq
\partial_{r_\alpha}\hat g_{\alpha\beta}(\r,\r')|_{\r'=\r}
=(1/2)\partial_{r_\alpha}\hat g_{\alpha\beta}(\r).
\label{C2}
\eeq
Substituting into Eq. (\ref{C1})
with the approximation (valid to the order considered) 
$\hat g_{\alpha\beta}(\bar\r,\r(\tau))\simeq \hat g_{\alpha\beta}(\bar\r)$,
leads immediately to Eq. (\ref{<f1>}).
The calculation of $\langle f_3[\hat\u]|\r(0)$=$\bar\r\rangle$, that leads to Eq. (\ref{<f3>}),
is completely analogous.

The calculation of $\langle f_2[\hat\u]|\r(0)$=$\bar\r\rangle$ is slightly more
involved and requires considering the correction terms in the response function,
Eq. (\ref{response1}). The starting point is the equation
\begin{eqnarray}
\langle\hat u_\alpha(\bar\r,\tau)\hat u_\beta(\bar\r,\tau')\delta(\r(0)-\bar\r)\rangle
=4\psi(\tau)\psi(\tau')
\nonumber
\\
\times
\langle\hat g_{\alpha\gamma}(\bar\r,\r(\tau)
\hat g_{\beta\eta}(\bar\r,\r(\tau'))\bar\partial_\gamma\bar\partial_\eta
\delta(\r(0)-\bar\r)\rangle
\nonumber
\\
-2\langle \hat u_\alpha(\r(\tau),\tau)\hat g_{\beta\gamma}(\bar\r,\r(\tau'))
R^\smalun_{\gamma\phi}(0;\r(\tau'),\tau')
\nonumber
\\
\times\bar\partial_\phi\delta(\r(0)-\bar\r)\rangle.
\label{C3}
\end{eqnarray}
For $\tau>\tau'$, 
the product $\hat u_\alpha(\r(\tau),\tau)R^\smalun_{\eta\phi}(0,\r(\tau'),\tau')$ in the 
second term to RHS of Eq. (\ref{C3}), from Eq. (\ref{response1}),
leads to a factor $2\psi(\tau'-\tau)\psi(\tau)
\partial_{r_\gamma(\tau)}\hat g_{\alpha\eta}(\bar\r,\r(\tau))=
2\psi(\tau'-\tau)\psi(\tau)
\bar\partial_\gamma\hat g_{\alpha\eta}(\bar\r,\r(\tau))$.
Substituting into Eq. (\ref{C3}), approximating
$\hat g_{\alpha\beta}(\bar\r,\r(\tau))\simeq\hat g_{\alpha\beta}(\bar\r,\r(\tau'))\simeq 
\hat g_{\alpha\beta}(\bar\r)$
and using Eq. (\ref{C2}), leads, after
little algebra, to Eq. (\ref{<f2>}).


\begin{thebibliography}{10}
\bibitem{nota1} For a given realization, 
the two velocity differences $\hat\u(\r,t)=u(\x+\r,t)-\u(\x,t)$ 
and  $\hat\u(\r',t')=u(\x+\r',t')-\u(\x,t')$ are defined at the same point $\x$
(the choice of $\x$ is part of the realization selection).
\end{thebibliography}
\end{document}